\documentclass[12pt]{article}
\usepackage{amsfonts,amsthm,amsmath,amssymb,upgreek}
\usepackage{hyperref}
\usepackage{cite}
\usepackage[paper=letterpaper,margin=1in]{geometry}
\usepackage{graphicx}
\usepackage{units}
\usepackage{color}
\usepackage[export]{adjustbox}
\usepackage{mathtools}

\usepackage{mciteplus}

\def\Tr{{\rm Tr }}
\newcommand{\CJ}{\mathcal{J}}
\newcommand{\CO}{\mathcal{O}}
\newcommand{\be}{\begin{equation}}
\newcommand{\ee}{\end{equation}}
\renewcommand{\hat}{\widehat}

\numberwithin{equation}{section}

\renewcommand{\title}[1]{\vbox{\center\bf{\Large{#1}}}\vspace{5mm}}
\renewcommand{\author}[1]{\vbox{\center#1}\vspace{5mm}}
\newcommand{\address}[1]{\vbox{\center\em#1}}
\newcommand{\email}[1]{\vbox{\center\tt#1}\vspace{5mm}}

\begin{document}

\begin{titlepage}

\begin{center}
\hfill \\
\hfill \\
\vskip .5cm

\title{Operator growth in the SYK model}

\author{Daniel A. Roberts,$^{ab}$ Douglas Stanford,$^a$  and Alexandre Streicher$^{cd}$}
\address{$^{a}$ School of Natural Sciences, Institute for Advanced Study,\\ Princeton, NJ 08540, USA

\vspace{10pt}

$^{b}$ Facebook AI Research, Facebook,\\ New York, NY 10003, USA

\vspace{10pt}

$^{c}$ Department of Physics, University of California,\\ Santa Barbara, CA 93106, USA

\vspace{10pt}
$^{d}$ Stanford Institute for Theoretical Physics, Stanford University,\\ Stanford, CA 94305, USA
}

\email{roberts@ias.edu | danr@fb.com, stanford@ias.edu, alex@physics.ucsb.edu}

\end{center}

\begin{abstract}
We discuss the probability distribution for the ``size'' of a time-evolving operator in the SYK model. Scrambling is related to the fact that as time passes, the distribution shifts towards larger operators. Initially, the rate is exponential and determined by the infinite-temperature chaos exponent. We evaluate the size distribution numerically for $N = 30$, and show how to compute it in the large-$N$ theory using the dressed fermion propagator. We then evaluate the distribution explicitly at leading nontrivial order in the large-$q$ expansion.
\end{abstract}

\end{titlepage}


\tableofcontents

\section{Introduction}

In quantum many-body systems, the butterfly effect is roughly the statement that time evolution takes simple (few-body) operators to complicated ones (many-body). This makes it possible for the disturbance of a single particle far in the past to have significant effects on all particles at a later time. In systems with spatial locality, this takes a while, since the disturbance has to spread through the system \cite{Lieb:1972wy,hastings2010locality,Roberts:2014isa,Aleiner:2016eni,Roberts:2016wdl,Gu:2016oyy,Chowdhury:2017jzb,Patel:2017vfp,Werman:2017abn,vonKeyserlingk:2017dyr,Nahum:2017yvy,Xu:2018xfz}. In nonlocal systems the process can be much faster. However, the concept of operator growth still makes sense if each term in the Hamiltonian only couples together a few degrees of freedom at a time. In this setting, simple operators still take time to become complicated \cite{dankert2009exact,Hayden:2007cs,Sekino:2008he,brown2010convergence,Brown:2012gy}.

A rough diagnostic of this effect is the commutator-squared between $W(t) = e^{iHt}We^{-iHt}$ and $V$, where $W,V$ are simple operators \cite{Larkin:1969abc,Almheiri:2013hfa,Shenker:2013pqa,Shenker:2013yza,Kitaev:2014t1}. The idea is that as time advances, $W(t)$ grows in such a way that it has nontrivial effects at almost any site in the system. As a result, it then fails to commute with other simple operators, such as $V$, and so $\langle [W(t),V]^2\rangle$ becomes order one. In the case where $W,V$ are fermionic operators, then one considers the anticommutator-squared instead.

In this paper we will consider another diagnostic, which is to compute the full probability distribution for the size of the time-evolving operator \cite{Roberts:2014isa}. To define this, one expresses $W(t)$ in a basis of operators organized by the number of ``simple'' operators that appear in a given product (the ``size''). Let's explain this more concretely for the case of the SYK model \cite{Sachdev:1992fk,Kitaev:2014t2}. In that case it is natural to take the simple operators to be the individual fermions, $\psi_i$. We choose $W$ to be one of those fermions, say $W = \psi_1$. The time-evolving $W(t)$ can be expanded as
\be\label{expansion}
\psi_1(t) = \sum_{s,\,a_1<...<a_s}2^{\frac{s-1}{2}}c_{a_1...a_s}(t)\psi_{a_1}...\psi_{a_s},
\ee
where $s$ is the ``size'' of the basis element, i.e. the number of elementary fermions that appear in the product. The factor $2^{\frac{s-1}{2}}$ is to compensate for the fact that we normalize the fermions so that $\psi^2 = \frac{1}{2}$. The probability distribution for size $s$ is then defined as
\be\label{sizeDef}
P_s(t) = \sum_{a_1<...<a_s}|c_{a_1...a_s}(t)|^2.
\ee
As time passes, this distribution shifts towards operators of larger size---the operator grows.

We can think of the $c_{a_1...a_s}(t)$ coefficients as describing a quantum wave function for the evolving operator. As we will see, in the infinite $N$ SYK model, this ``operator wave function'' can be understood as a standard wave function for a quantum particle moving on a special, rapidly expanding graph shown in Fig.~\ref{fig:graph}a. With time evolution, most of the particle's wave function moves deeper into the graph at an exponentially growing rate. This corresponds to the operator becoming larger and more complicated. 

The fact that the rate is exponential is because the graph on which the particle is moving becomes more highly connected as we move deeper. In terms of the growing operator, this reflects the fact that once an operator has already become quite large, it has many different ways to grow larger still. This is the basic origin of exponential early-time behavior of correlators such as $\langle [W(t),V]^2\rangle$, diagnosed by the chaos exponent (or many-body quantum Lyapunov exponent) $\lambda_L$ \cite{Kitaev:2014t1,Maldacena:2015waa}.

Although most of the wave function moves rapidly into the graph, there is a small exponentially decaying tail for the particle to remain at (or return to) the root of the graph. In operator language this is the amplitude for $\psi_1(t)$ to remain equal to $\psi_1(0)=\psi_1$, or more explicitly $\frac{1}{2^{N/2}}\Tr[\psi_1(t)\psi_1(0)]$, the infinite temperature two point function. This correlator exponentially decays because most of the wave function is leaking into the space of complicated many-fermion operators.

It is a challenging problem to go beyond this qualitative discussion and actually compute $P_s(t)$. In the rest of the paper, we will discuss some partial results for this quantity, mainly in the large $N$ theory. In particular, we will explain an equivalent particle-moving-on-a-graph problem, and we will show how to sum the `melonic' (infinite $N$) perturbation theory for the wave function using the dressed infinite-temperature two-point function of fermions. However, the dressed two-point function is not known analytically except for the large-$q$ SYK model. So the only place where we will succeed in computing the operator wave function is at infinite $N$ and at leading order in large $q$. We will use this wave function to compute a few things, including the (previously known) infinite-temperature chaos exponent. 

Before we begin with the main calculation, we will make three preliminary comments.

\section{Preliminary comments}
\subsection{A classical model}\label{appendixB}
Before analyzing the quantum problem, we can discuss an analog of operator growth for a classical model of many-body chaos. This model was previously considered in the context of high energy scattering in weakly coupled gauge theory \cite{mueller1995unitarity}. It goes like this. Suppose we have a collection of $N$ particles where we initially label one as infected and the others as healthy. The rule for time evolution is that with some probability $\gamma$ per unit time, any infected particle can heal itself at the cost of infecting $(q{-}1)$ random other particles. For simplicity, we assume that the total number $N$ is large enough and the time is short enough that the infected particles are always very dilute. If we are interested in the probability that some randomly chosen particle will be infected after time $t$, we can proceed in three different ways. The three ways get increasingly complicated, but each gives an interesting perspective.

The first way is to simply notice that the expected number of infected particles $\langle N_{inf}\rangle$ is growing according to $\frac{d \langle N_{inf}\rangle}{dt} = (q-2)\gamma \
\langle N_{inf}\rangle$. This leads to $\langle N_{inf}\rangle = e^{(q-2)\gamma t}$. The probability that a randomly chosen particle will be infected at time $t$ is simply $\frac{\langle N_{inf}\rangle}{N} = \frac{1}{N}e^{(q-2)\gamma t}$. This type of intuition was used in early discussions of scrambling by quantum circuits \cite{Sekino:2008he}, and it is related to the kinetic equation method used for the OTOC in weakly coupled theories in \cite{Aleiner:2016eni}.

A second way is to follow the possible chains of events that lead to our random final particle getting infected, and add up all of the probabilities. To start, the simplest way it could happen is that the original infected particle never infects any other particles, but by chance it happens to be our randomly selected particle at time $t$. The probability for this is $\frac{1}{N}e^{-\gamma t}$. The second simplest thing would be for the original particle to infect one set of $(q{-}1)$ particles, for one of these to be our randomly chosen particle, and for this one to not infect any further particles. The probability for this is
\be
\frac{q-1}{N}\, \gamma \int_0^t dt' e^{-\gamma t'} e^{-\gamma (t-t')} =  \frac{q-1}{N}\, \gamma t \, e^{-\gamma t}.
\ee
Summing over all possible chains of infection, we find the probability
\be
\frac{1}{N}\sum_{k = 0}^\infty \frac{\big((q{-}1)\, \gamma t\big)^k}{k!}e^{-\gamma t} = \frac{1}{N}e^{(q-2)\gamma t}.
\ee
This method of calculation is very similar to the calculation of the OTOC by ladder diagrams, where $k$ labels the number of rungs in the ladder (see \cite{kitaevfirsttalk,Kitaev:2014t2} in the SYK model and \cite{Shenker:2014cwa,Stanford:2015owe,Chowdhury:2017jzb,Patel:2017vfp,Werman:2017abn} in weakly coupled theories). In particular the crucial factors of $(q-1)$ that appear here and that also appear in the ladder diagrams. These factors correct for the fact that we are only following one possible chain of infection, picking one out of the $(q{-}1)$ particles infected in each given event.

A final way, analogous to the discussion of the quantum problem in this paper, is to calculate the probability distribution for the number of infected particles at time $t$, and then take the expectation value explicitly. To do this we can use the equation
\be
\frac{dP(s_k,t)}{dt} = \gamma s_{k-1} \, P(s_{k-1},t) - \gamma s_k \, P(s_k,t), \hspace{20pt} s_k = 1 + (q-2)k.
\ee
The solution with initial conditions $P(s_k,0) = \delta_{k,0}$ is
\be
P(s_k,t) = \frac{\Gamma(k+\frac{1}{q-2})}{\Gamma(k+1)\Gamma(\frac{1}{q-2})}e^{-\gamma t}\big(1 - e^{-(q{-}2)\gamma t}\big)^k.
\ee
With this probability distribution, we can of course find the same answer as with the other two methods for the probability that a random particle is infected, by taking the sum $\sum_{k} \frac{s_k}{N}P(s_k,t) = \frac{1}{N}e^{(q-2)\gamma t}$. It is interesting to note that if we scale time so that $\gamma \sim \frac{1}{q}$, this formula is quite similar to the probablity distribution we will find in the SYK model at large $q$, see (\ref{resummed}).

\subsection{A numerical plot for $N = 30$}
We would now like to show an example numerical plot of $P_s(t)$ for the quantum problem. To begin we should explain how this can be computed numerically. The SYK model with $N$ Majorana fermions lives in a Hilbert space of dimension $2^{N/2}$. The space of operators acting on that Hilbert space can be understood as a Hilbert space in its own right, with inner product given by
\be\label{innerprod}
\left(A,B\right) \equiv \frac{1}{2^{N/2}}\Tr\left[A^\dag B\right].
\ee
In this ``operator Hilbert space'' we can decompose $\psi_1(t)$ in a basis of operators of definite size, as in (\ref{expansion}). A formula equivalent to (\ref{sizeDef}) is
\be\label{sizeDef2}
P_s(t) \equiv \sum_{\mathcal{O}\in \{\text{op.s of size }s\}}\frac{\left|\big(\mathcal{O},\psi_1(t)\big)\right|^2}{\left(\mathcal{O},\mathcal{O}\right)\left(\psi_1,\psi_1\right)}.
\ee
Here, the sum is over an orthogonal basis $s$-fermion operators, for example all ${N\choose s}$ operators of the form $\psi_{a_1}\dots \psi_{a_s}$ with $a_1<...<a_s$.

\begin{figure}[ht]
\begin{center}
\includegraphics[width = .9\textwidth]{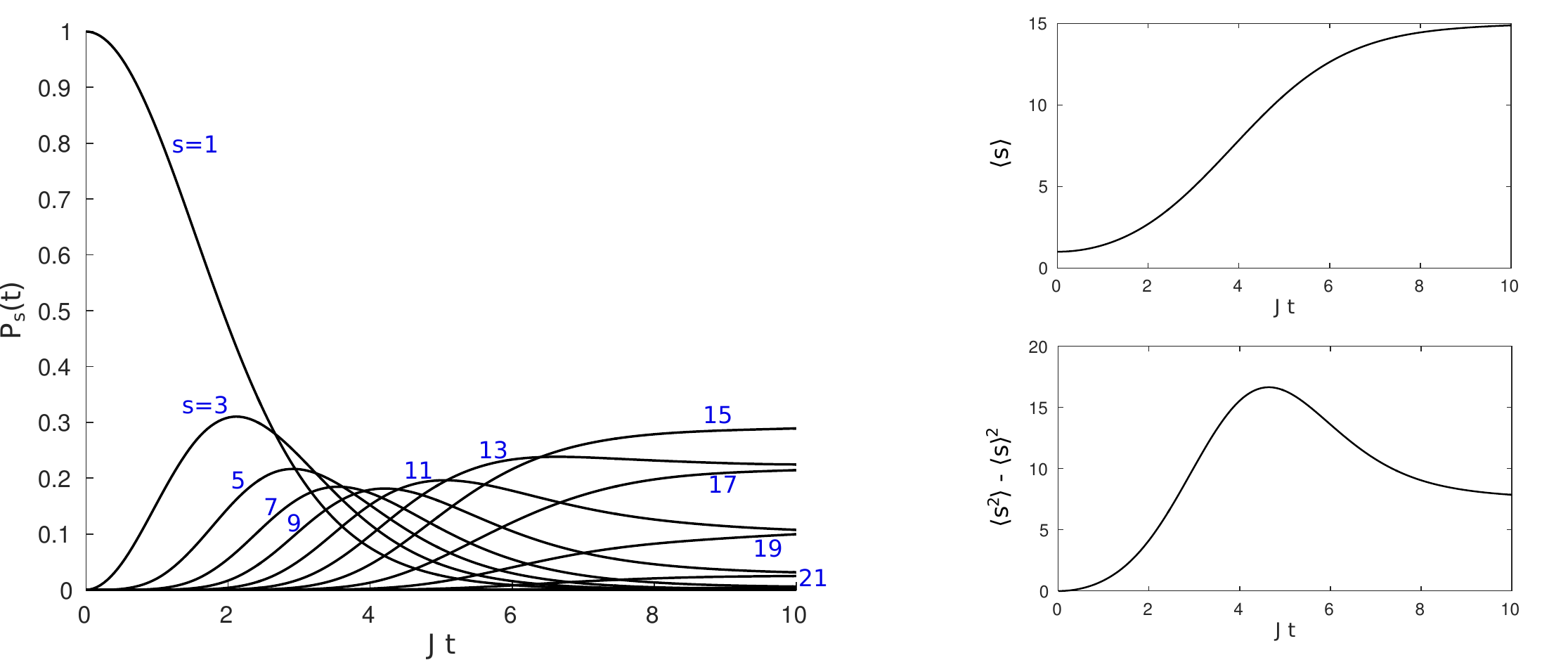}
\caption{{\small Exact diagonalization numerics for SYK with $N = 30$ and $q = 4$ (see \S~\ref{sec:graph} for the definition of this model). At left, we plot the distribution of sizes in the operator $\psi_1(t)$ as a function of time. Notice that in the early phase, the peaks occur more rapidly as time passes. This is because already-large operators can grow faster than small ones. The ``scrambling time'' where the operator reaches full size would fall somewhere around three-quarters of the way through the plot. At right, we plot both the mean value and the variance of the size.}}\label{fig:plotN=30}
\end{center}
\end{figure}

To evaluate this in practice for reasonably small values of $N$, we can compute $\psi_1(t)$ by exact diagonalization and exponentiation of the Hamiltonian and then evaluate the sum over operators in (\ref{sizeDef2}) by random sampling. In Fig.~\ref{fig:plotN=30} we show the result of this computation. We also plot the expected value and variance of the size as a function of time.
At $t = 0$ all of the probability is concentrated in size one, but as time passes we see successive peaks in larger sizes as the probability mass moves towards larger operators. At late time, the distribution appears to converge to the size distribution of a random fermionic operator. For such an operator, $P_s$ is proportional to the total number of operators of size $s$ (for $s$ odd only), which gives $P_s \rightarrow {N\choose s}2^{1-N}$. So for example, the most common size at late time is $\frac{N}{2} = 15$.

\subsection{The size and infinite-temperature OTOCs}\label{sec:size-and-com}
As a final preliminary comment, we would like to show that out-of-time-order correlators at infinite temperature are related to the expectation value of $s$ in the distribution $P_s(t)$. We define $A(t)$ as the typical anticommutator squared at infinite temperature between $\psi_1(t)$ and a single-fermion operator
\be
A(t) \equiv \frac{1}{N}\sum_j \frac{1}{2^{N/2}}\text{Tr}\left[\{\psi_1(t),\psi_j\}^2\right] = \frac{1}{N}\sum_{j}\Big(\{\psi_1(t),\psi_j\},\{\psi_1(t),\psi_j\}\Big).
\ee
Here, we have averaged over the index of the second operator $\psi_j$. The growth of this object is a useful diagnostic for quantum chaos and is simply related to other infinite-temperature out-of-time-order correlators. 

Inserting the expansion of $\psi_1(t)$ in (\ref{expansion}), we have
\begin{align}
A(t) &= \frac{1}{N}\sum_j \sum_{\substack{s,\,a_1<...<a_s\\s',\,b_1<...<b_{s'}}}2^{\frac{s+s'-2}{2}}c_{a_1...a_s}^*(t)\,c_{b_1 ... b_{s'}  }(t)\, \Big( \{\psi_{a_1} ... \psi_{a_s}, \psi_j\}, \{\psi_{b_1} ... \psi_{b_{s'}}, \psi_j\} \Big).
\end{align}
In order to simplify this expression, we can use that the $\psi_{a_1}...\psi_{a_s}$ operators are orthogonal with respect to our inner product $(\cdot,\cdot)$ and that this is preserved after taking anticommutators with $\psi_j$. In fact, when $a_1<...<a_s$ and $b_1<...<b_{s'}$, one has the useful formula
\be
\Big( \{\psi_{a_1} ... \psi_{a_s}, \psi_j\}, \{\psi_{b_1} ... \psi_{b_{s'}}, \psi_j\} \Big)   
= \Bigg\{
\begin{matrix*}[l]
2^{1-s}, &   \{a_1... a_s\} = \{b_1... b_{s'}\} \ \text{  and  }  \ j \in \{a_1 ... a_s\}, \\
0, & \text{ else}.
\end{matrix*}\label{eq:average-over-j}
\ee
This formula collapses the sum over $\{a_1...a_s\}$ and $\{b_1...b_{s'}\}$ to the diagonal terms. It also allows us to sum over $j$, getting a factor of $s$ from the $s$ different values of $j$ for which we get a nonzero answer. We find
\be
A(t) = \frac{1}{N}\sum_{s,\, a_1<...<a_s}s \, |c_{a_1...a_s}(t)|^2 = \frac{1}{N}\sum_{s}s\, P_s(t) = \frac{\langle s\rangle}{N}.\label{anticommutator-squared}
\ee
In other words, $A(t)$ is simply related to the mean size of the operator $\psi_1(t)$.

\section{The graph of operators}\label{sec:graph}
We will now proceed with the main part of the paper. From this point forward we will be discussing the SYK model in the large $N$ limit. Our conventions for SYK \cite{Sachdev:1992fk,Kitaev:2014t2} are that the Hamiltonian is
\be
H = i^{q/2}\sum_{1\le a_1<...<a_q\le N}J_{a_1 ... a_q}\psi_{a_1} ... \psi_{a_q}, \hspace{20pt}\{\psi_a,\psi_b\} = \delta_{ab}.
\ee
Here $J_{a_1 ... a_q}$ is an antisymmetric tensor, drawn from a Gaussian distribution, with mean zero and the property that the square of a given component has the average value
\be
\langle J_{a_1 ... a_q}^2\rangle = \frac{(q{-}1)!}{N^{q-1}}J^2, \qquad \text{(no sum)}.\label{eq:J-varianace}
\ee
Here, we introduced the dimensionful constant $J$. We will also use $\mathcal{J}$, which differs by a $q$-dependent factor as in \cite{Maldacena:2016hyu}
\be
J^2 = \frac{2^{q-1}}{q}\mathcal{J}^2.
\ee

We would like to understand the time evolution of a particular fermion operator $\psi_1(t)$ in the large $N$ limit of this model. A key simplification will be that the this time evolution stays within a particular class of operators, which consist of many fermions contracted together in various ways with the $J_{a_1\dots a_q}$ tensor.

It is convenient to organize this class of operators by their size, which as always refers to the number of elementary fermion operators. We will sometimes use ``generation'' in place of size, where generation refers to the number of times we have to commute $H$ with $\psi_1$ for the operator to first appear in the Baker-Campbell-Hausdorff series for $\psi_1(t)$ (for further discussion of this perspective, see \S~\ref{sec:discussion}). Size $s$ and generation $k$ are related by
\be
s = 1 + (q-2)k, \hspace{20pt} s = \text{size}, \hspace{20pt} k = \text{generation}.
\ee
Let us now discuss the types of operators that appear in the time evolution of $\psi_1(t)$, choosing $q = 4$ for simplicity.
\begin{itemize}
\item {\bf Generation zero:} At infinite $N$, the only size-one operator that appears in $\psi_1(t)$ is simply $\psi_1$ itself. It will be convenient to work with operators that are orthonormal with respect to the inner product defined in (\ref{innerprod}). A normalized version of the operator $\psi_1$ is simply $\sqrt{2}\psi_1$, which we denote as
\be
\includegraphics[scale = 2,valign=c]{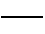} \   = \mathcal{O}_0 = 2^\frac{1}{2}\psi_1.
\ee
Our notation for this operator as a horizontal line will become clear from further examples below.

\item {\bf Generation one:} At generation one (size $q-1$), the operator that appears in the time evolution is simply the commutator of the Hamiltonian with $\psi_1$. The normalized version of this operator is
\be
\includegraphics[scale = 2,valign=c]{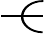} \   = \mathcal{O}_1 = 2^{\frac{3}{2}}\sum_{a<b<c}\frac{J_{1abc}}{J}\psi_a\psi_b\psi_c.
\ee
We can interpret this operator as follows. The original fermion $\psi_1$ has split into $q-1$ fermions by a single action of the Hamiltonian.

\item {\bf Generation two:} In the second generation, it will be convenient to divide the operator into three (more generally $q-1$) terms, corresponding to a further division into $q-1$ fermions of any of the fermions present in the operator $O_1$. These distinct terms correspond to the operators
\begin{align}
\includegraphics[scale = 2,valign=c]{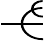} \   &= \mathcal{O}_2^{(1)} = 2^\frac{5}{2}\sum_{\substack{a_1<a_2<a_3 \\ b_1<b_2<b_3}}\frac{J_{1a_1a_2a_3}J_{a_1b_1b_2b_3}}{J^2}\psi_{a_2}\psi_{a_3}\psi_{b_1}\psi_{b_2}\psi_{b_3},\label{eq:operator-k-2-ell-1} \\
&\vdots\notag\\
\includegraphics[scale = 2,valign=c]{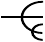} \   &= \mathcal{O}_2^{(3)} = 2^\frac{5}{2}\sum_{\substack{a_1<a_2<a_3 \\ b_1<b_2<b_3}}\frac{J_{1a_1a_2a_3}J_{a_3b_1b_2b_3}}{J^2}\psi_{a_1}\psi_{a_2}\psi_{b_1}\psi_{b_2}\psi_{b_3}.
\end{align}
Our notation with the fan diagrams is that the three daughter lines coming out of a vertex are always ordered such that the index of the top line is less than the index of the middle line, which is less than the index of the bottom line. Because of this ordering convention, the operators shown above are different from each other. 
\item {\bf Generation three:} In the third generation, there are different kind of operators that can appear, corresponding to the division of a fermion that was ``born'' in the first generation or the second generation. For example, we have the operators
\be
\includegraphics[scale = 2,valign=c]{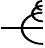} \hspace{20pt}\dots\hspace{20pt}\includegraphics[scale = 2,valign=c]{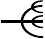}\hspace{20pt} \dots
\ee
\item {\bf Generation $k+1$:} More generally, the operators for generation $k+1$ are obtained by considering all of the operators at generation $k$, and for each one, allowing one of the fermions to divide further, contracting with a $J_{a_1\dots a_q}$ symbol, and normalizing. 
Graphically, we simply turn one of the lines into a fan. 
\end{itemize}
In the infinite $N$ limit, these operators all have definite size and are orthogonal. Note that at finite $N$, some of the indices of the fermions might happen to be the same. Using that $\psi_a^2 = \frac{1}{2}$, this would imply that the operator is actually of smaller size than $1 + (q-2)k$. However, this does not happen at infinite $N$.

Now that we have discussed the set of operators that we will use, we can describe the evolution of $\psi_1(t)$. The idea is that the operators we have described form a graph, and the time evolution of the operator is simply the quantum evolution of a particle moving on the graph. More precisely, we can think of the space of operators being a Hilbert space with inner product (\ref{innerprod}). In the infinite $N$ SYK model, the operators we dissused above correspond to an orthonormal basis for a subspace of the space of all possible operators. It is helpful to think about these operators $\mathcal{O}_k^{(\ell)}$ as basis states $|\mathcal{O}_k^{(\ell)}\rangle$ for an abstract particle that represents the evolving operator $\psi_1(t)$. The Heisenberg equation $\frac{d}{dt}\psi_1(t) = i[H,\psi_1(t)]$ is an ordinary Schrodinger equation acting in this space, for an appropriate Hamiltonian $\hat{H}$:
\be
\frac{d}{dt}|\psi_1(t)\rangle = -i\hat{H}|\psi_1(t)\rangle, \hspace{20pt} \langle A|\hat{H}|B\rangle \equiv -\frac{1}{2^{N/2}}\Tr\left(A^\dagger[H,B]\right).
\ee

\begin{figure}[ht]
\begin{center}
\includegraphics[width = \textwidth]{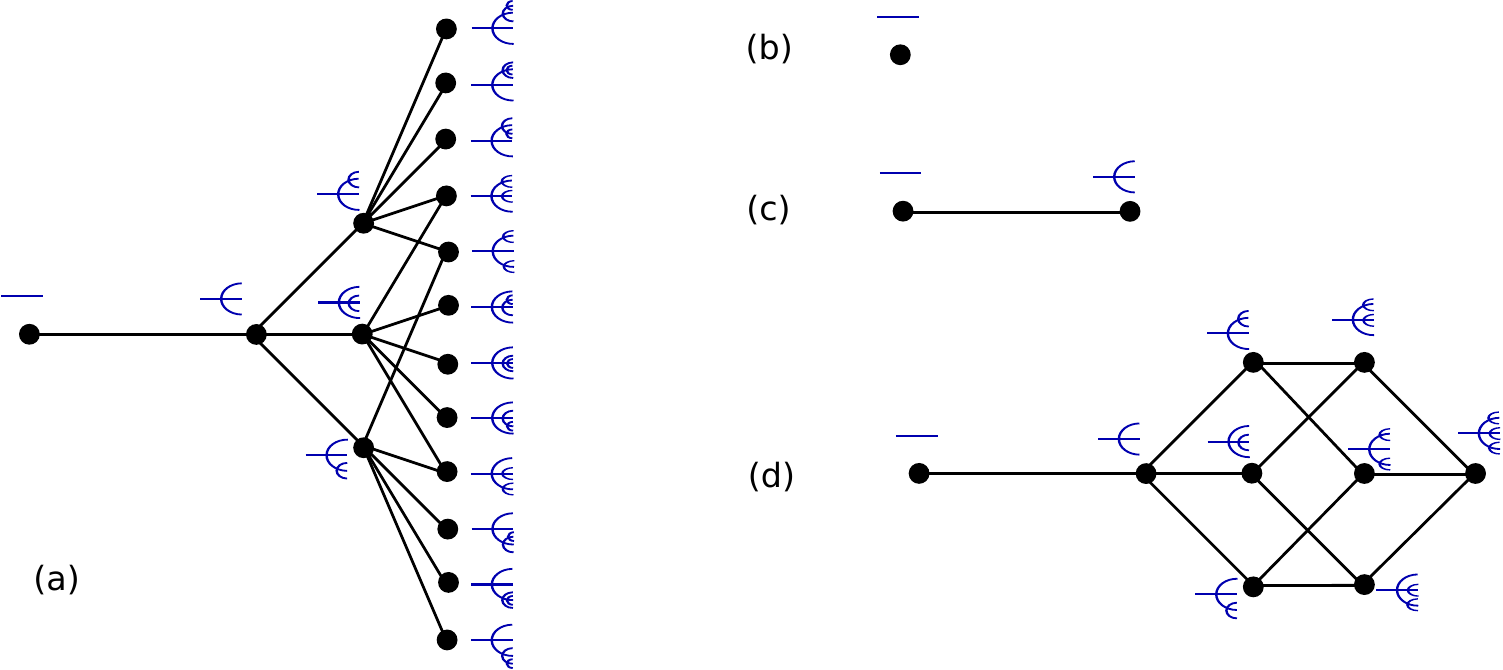}
\caption{{\small The graph of operators. In (a) we show the first four layers. Vertices correspond to basis operators, whose associated fan diagrams are indicated in blue. The problem of the time evolution of $\psi_1(t)$ in the large $N$ theory is equivalent to the motion of a quantum particle on this graph (extended to further layers). In (b), (c), and (d), we show versions of the graph where we limit the recursive depth of the fan diagrams. The return amplitude on these graphs gives the zeroth, first and second iterations of the real-time Schwinger-Dyson equations. For any finite cutoff these amplitudes oscillate in time, but for the infinite graph the return amplitude decays exponentially.}}\label{fig:graph}
\end{center}
\end{figure}

We can now explain the point of the basis of operators that we have chosen. The nice feature is that in this basis, $\hat{H}$ is proportional to the adjacency matrix on the graph,
\be
\hat{H} = 2^{1-\frac{q}{2}}J\cdot (\text{adjacency matrix}).
\ee
The adjacency matrix is defined as the matrix that has a $1$ at location $i,j$ if $i,j$ are vertices connected by an edge, and a zero otherwise. So, for example, we have
\be
\langle \includegraphics[scale = 1.3,valign=c]{gen0.pdf}|\hat H|\includegraphics[scale = 1.3,valign=c]{gen1.pdf}\rangle  =  2^{1-\frac{q}{2}}J, \hspace{20pt}\langle \includegraphics[scale = 1.3,valign=c]{gen1.pdf}|\hat H|\includegraphics[scale = 1.2,valign=c]{gen2.pdf}\rangle  =  2^{1-\frac{q}{2}}J, \hspace{20pt} \langle \includegraphics[scale = 1.3,valign=c]{gen0.pdf}|\hat H|\includegraphics[scale = 1.2,valign=c]{gen2.pdf}\rangle  = 0.
\ee
The evolution of the operator $\psi_1(t)$ in the large $N$ theory is therefore simply the quantum evolution of a particle moving on the graph shown in Fig.~\ref{fig:graph}, with initial condition that the particle starts out at the leftmost vertex.

An obvious feature of this graph is that it is rapidly expanding. The degree of a vertex (the number of neighbors) grows roughly linearly with generation $k$. This corresponds to the fact that at generation $k$, the operators contain $1+(q-2)k$ fermions, and the Hamiltonian can act on any one of those, turning a single fermion into a fan of fermions and producing a new operator at generation $k+1$. The fact that the degree is growing with distance in this way means that this graph expands more rapidly than e.g. a Cayley graph/Bethe lattice/discretization of hyperbolic space, for which the degree is constant. 

We would like to call attention to two qualitative features of the evolution of a particle on such a graph.

\begin{enumerate}
\item A basic effect is that particles tend to move to the right, towards more complicated operators. This is because at any given vertex, there tend to be many more edges leading to the right than to the left: there are more ways for the operator to grow than to shrink. We expect this to lead to exponential decay of the amplitude that the particle remains (or returns) to the original leftmost vertex $\psi_1$. This amplitude is simply the correlation function $\frac{2}{2^{N/2}}\Tr\left(\psi_1(t)\psi_1(0)\right)$. It exponentially decays due to the wave function for $\psi_1(t)$ leaking more and more into the space of complicated operators orthogonal to $\psi_1(0) = \psi_1$.

The graph picture gives an intuitive explanation for why the real-time correlator should exponentially decay, but it does not give an efficient method for computing the decay rate. The best way we know of to compute the correlator is by numerically solving the Schwinger Dyson equations in real time to compute the retarded propagator. At infinite temperature, the real-time equations are simply given by
\be\label{SD}
G_R(\omega) = \frac{1}{-i\omega + \Sigma(\omega)+\epsilon}, \hspace{20pt} \Sigma(t) = 2^{2-q} J^2 G_R(t)^{q-1}, \hspace{20pt} f(t) = \int\frac{d\omega}{2\pi}f(\omega)e^{-i\omega t}.
\ee
At infinite temperature, the retarded propagator is simply $G_R(t) =\frac{ 2\theta(t) }{2^{N/2}}\text{Tr}\left(\psi_1(t)\psi_1(0)\right)$. For $t>0$ this is exactly the return amplitude for the quantum particle to be at the leftmost vertex of the graph as a function of time. In the next section, we will see how to use the solution to these equations to write a formula for the wave function on other vertices. For now, we will make a side comment. One way to solve these SD equations is to start with the free answer $\Sigma = 0$ and simply iterate the equations. The function $G_R(t)$ that we get after a finite number of iterations sums a set of diagrams where we cut off the recursive structure of the melon diagrams at some level. For example, after iterating zero times, we simply take the free propagator. After one iteration or two iterations, respectively, we are effectively summing diagrams of the form
\be\label{diags}
\includegraphics[scale = 1.3,valign=c]{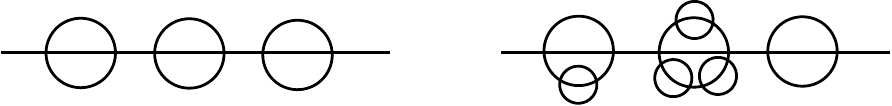}
\ee
where all lines represent free propagators. Summing these diagrams is equivalent to evaluating the return amplitude for a particular cutoff version of the graph, where we keep all vertices that correspond to fan diagrams with `recursive depth' equal to or less than the number of iterations of the SD equations. For example, the cutoffs corresponding to the zero-th, first and second iterations of the SD equations are shown in Fig.~\ref{fig:graph} in panels (b), (c) and (d).

This gives a perspective on why we get exponential decay of the two point function $\frac{2}{2^{N/2}}\Tr\left(\psi_1(t)\psi_1(0)\right)$. For example, consider the self-energy diagrams on the left in (\ref{diags}). These describe oscillation between the operator $\psi_1$ and $J_{1abc}\psi_a\psi_b\psi_c$. In the graph picture, it represents a particle that is moving between the two states of the simple graph shown in (c) of Fig.~\ref{fig:graph}. The result is a return amplitude that oscillates in time, $\cos(2^{1-\frac{q}{2}} J t)$. If we consider the SD equations after two iterations, we are studying a particle moving in the somewhat more complicated graph shown in (d). It still oscillates, but the return amplitude has a somewhat lower average value. For any finite cutoff, or any finite iteration of the SD equations, we will get a correlator that oscillates in time. But in the limit where we study the infinite graph, the return amplitude decays exponentially because the particle can continue moving to the right forever in the infinite graph

\item Another important qualitative feature is that the expected size of the operator grows exponentially in time. This is because the degree of the graph is growing linearly with the generation $k$. The timescale for evolution from generation $k$ to $k+1$ is proportional to the inverse of the degree, which is proportional to $1/k$. So as the particle moves farther out into the graph, it speeds up proportionally to its distance. This leads to the expectation value of $k$ growing exponentially with time.\footnote{To make this argument more reliably, one needs to know that the number of vertices at generation $k$ is growing only exponentially in $k$, and not faster. The precise formula for the number of vertices at generation $k$ is $\frac{1}{k}{ k(q-1)\choose k-1}$, which grows exponentially in $k$ for large $k$.}
\end{enumerate}

\section{Computing the wave function on the graph}
In principle, one could evaluate the wave function for the particle moving on the graph by directly studying that problem. However, it is more convenient to translate the problem into a correlation function in the infinite temperature SYK model and then re-sum the `melonic' SYK perturbation theory in the usual way.

Let's imagine that we want to compute the wave function that corresponds to the time evolution of the operator $\mathcal{O}_0(t) = 2^{\frac{1}{2}}\psi_1(t)$. We can write this explicitly as
\be\label{timecontour}
\langle \mathcal{O}_k^{(\ell)}|e^{-i\hat{H}t}|\mathcal{O}_0\rangle = \frac{2^{\frac{1}{2}}}{2^{N/2}}\Tr\left(\mathcal{O}_k^{(\ell)} \ \psi_1(t)\right)=\includegraphics[scale = 1.3,valign=c]{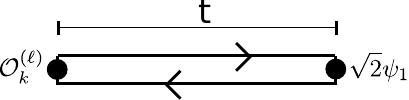}
\ee
The diagram in the last expression is the time contour for a path integral that evaluates the correlator. The two horizontal lines with arrows on them represent the forwards and backwards time evolution operators in the expression $\psi_1(t) = e^{iHt}\psi_1e^{-iHt}$. In order to evaluate this quantity by perturbation theory, we should integrate interaction vertices of the SYK model everywhere on this folded time contour, connecting the loose propagators either to the $\psi_1$ operator at the right end, or the fermions in whatever $\mathcal{O}_k^{(\ell)}$ basis operator we are considering.

The simplest case is the return amplitude, when we take $\mathcal{O}_k^{(\ell)} = \mathcal{O}_0 \equiv 2^{\frac{1}{2}}\psi_1$, which we also represent with the symbol $\includegraphics[scale = 1.1,valign=c]{gen0.pdf}$. Then the quantity we are computing is
\be
\langle \includegraphics[scale = 1.3,valign=c]{gen0.pdf}|e^{-i\hat{H}t}|\includegraphics[scale = 1.3,valign=c]{gen0.pdf}\rangle =\frac{(2^{\frac{1}{2}})^2}{2^{N/2}}\Tr\left(\psi_1(0)\psi_1(t)\right) = 2G(t),
\ee
namely twice the two point function at infinite temperature for time separation $t$. If we like, we can write this (for $t>0$) as the retarded propagator, since at infinite temperature $G_R(t) = \frac{1}{2^{N/2}}\Tr\left(\{\psi(t),\psi(0)\}\right)\theta(t) = 2\theta(t)G(t)$. So the answer for the return amplitude is given by the solution to the Schwinger-Dyson equations (\ref{SD}).

It is helpful to have a quick look at the perturbation theory that generates the SD equations. At large $N$, the perturbation theory for the return amplitude looks like the following
\begin{align}
\langle \includegraphics[scale = 1.3,valign=c]{gen0.pdf}|e^{-i\hat{H}t}|\includegraphics[scale = 1.3,valign=c]{gen0.pdf}\rangle &= \hspace{5pt}\includegraphics[scale = 1.3,valign=c]{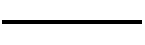}\hspace{10pt} + \includegraphics[scale = 1.3,valign=c]{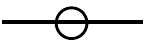}\hspace{10pt}\dots + \hspace{10pt} \includegraphics[scale = 1.3,valign=c]{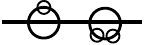}\hspace{10pt} + \dots  \\&=\hspace{5pt}\includegraphics[scale = 1.3,valign=c]{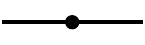}
\end{align}
Let us explain this notation. In the first line, in the Feynman diagrams, the two endpoints represent the operators $\sqrt{2}\psi_1$ inserted at time zero and time $t$. The lines in the Feynman diagrams represent free propagators, so the first diagram is simply $1 = (\sqrt{2})^2\cdot \frac{1}{2}$, where the two factors of $\sqrt{2}$ are for the normalizations of the external operators, and the $\frac{1}{2}$ is a free fermion propagator. When we have interaction diagrams, we need to take care to sum over whether the interaction vertex is on the ``forwards'' or ``backwards'' portion of the time contour. It is easy to check that these contributions cancel unless all vertices are ordered in time in the same way that they are ordered in the diagram. In this case, the contributions from the two portions of the contour add together, giving an extra factor of two for each vertex. So for example the second diagram gives $(\sqrt{2})^2\cdot (2iJ)^2\cdot \frac{t^2}{2}\cdot (\frac{1}{2})^5$. Here the $(2iJ)^2$ is for the two interaction vertices, the $\frac{t^2}{2}$ is for the integral over two ordered points between zero and $t$, and the $(\frac{1}{2})^5$ is for the five free fermion propagators. This evaluates to $-\frac{J^2t^2}{8}$. In the second line, we represent a dressed retarded propagator, which is equal to the return amplitude, as a line with a black dot in the middle.

The next simplest case is when we take $\mathcal{O}_k^{(\ell)} = \mathcal{O}_1 = \includegraphics[scale = 1.1,valign=c]{gen1.pdf}$. Now we need to evaluate a correlation function of a composite operator built out of three fermions, and the single-fermion operator $\psi_1(t)$. The lowest order diagram for this involves expanding down a single copy of the interaction vertex $J_{1abc}\psi_1\psi_a\psi_b\psi_c$, where the $\psi_1$ is contracted with our operator $\psi_1(t)$, and the other fermions are contracted with the $\mathcal{O}_1$ operator at time zero. Note that this Feynman diagram has the same structure as the fan diagram $\includegraphics[scale = 1.1,valign=c]{gen1.pdf}$ that we used to label the operator itself. At infinite $N$, the only other diagrams that contribute are `melonic' decorations of this diagram, as in
\be
\langle \includegraphics[scale = 1.3,valign=c]{gen1.pdf}|e^{-i\hat{H}t}|\includegraphics[scale = 1.3,valign=c]{gen0.pdf}\rangle = \hspace{5pt}\includegraphics[scale = 1.3,valign=c]{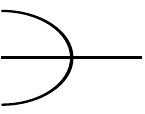}\hspace{10pt} + \dots + \hspace{10pt} \includegraphics[scale = 1.3,valign=c]{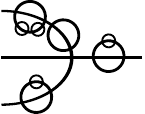}\hspace{10pt} + \dots 
\ee
These decorations can be summed by replacing the free propagators by dressed propagators that solve the Schwinger-Dyson equations. The full answer, including the numerical factor from the normalization of the operators, is
\begin{align}
\langle \includegraphics[scale = 1.3,valign=c]{gen1.pdf}|e^{-i\hat{H}t}|\includegraphics[scale = 1.3,valign=c]{gen0.pdf}\rangle &= -2^{1-\frac{q}{2}}iJ\int_0^t dt_1 G_R(t_1)^{q-1}G_R(t-t_1), 
\\
\langle \includegraphics[scale = 1.3,valign=c]{gen1.pdf}|e^{-i\hat{H}t}|\includegraphics[scale = 1.3,valign=c]{gen0.pdf}\rangle &=  -2^{1-\frac{q}{2}}iJ\int_0^t dt_1 \hspace{5pt}\includegraphics[scale = 1.3,valign=c]{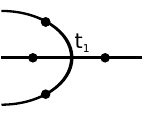}
\end{align}

This simple pattern persists for arbitrary operators $\mathcal{O}_k^{(\ell)}$: to compute the wave function, we can simply interpret the fan diagram of the operator $\mathcal{O}_k^{(\ell)}$ itself as a Feynman diagram, where all of the edges are dressed retarded propagators $G_R(t)$. We then integrate over the times of the vertices (subject to the ordering constraint which is imposed by the $\theta(t)$ in the retarded propagator). Including the correct numerical prefactor, one has for $k\ge 1$
\be
\langle \mathcal{O}_k^{(\ell)}|e^{-i\hat Ht}|\includegraphics[scale = 1.3,valign=c]{gen0.pdf}\rangle = ( -2^{1-\frac{q}{2}}iJ)^k\times \int dt_1\dots dt_k \Big[\text{$G_R$ factors reflecting fan diagram}\Big].\label{eq:expression-for-wave-function}
\ee

This gives an algorithm for computing the wave function of the particle moving on the graph, i.e. the time evolving operator. However there are two problems. First, in general we do not have an exact expression for the infinite temperature $G_R(t)$. Second, the number of different fan diagrams grows rapidly with generation $k$. In the special case of large $q$, both problems go away, because there is a known formula for $G_R(t)$, and as we will see, the different fan diagrams at a given generation are all proportional to the same function of time.

\section{The wave function in the large-$q$ SYK model}
In this section we will evaluate the wave function and corresponding probability distribution $P_s(t)$ to leading nontrivial order in the large-$q$ SYK model, namely $\frac{1}{q}$. To begin we will do a straightforward large $q$ analysis, where $t$ does not scale with $q$. This approximation breaks down at times of order $q$, and we will comment on how to resum $t/q$ effects at the end of the section.

At large $q$ and infinite temperature, there is a simple expression for the product of $q$ propagators \cite{Maldacena:2016hyu}
\be
G_R(t)^q = \frac{\theta(t)}{\cosh^{2} \CJ t} +O(1/q).\label{eq:q-retarded-propagators-large-q}
\ee
Taking the $1/q$-th power of this, we find that
\be
G_R(t) = \theta(t) + O(1/q),
\label{eq:retarded-propagator-large-q}
\ee
so a single propagator is almost given by the free answer. The fact that $G_R(t)^q$ is nontrivial will lead to an interesting wave function. However, the computation will be simplified by the fact that any fixed $O(1)$ number of propagators are simply step functions, which means that once the ordering of time arguments are imposed we can set them equal to unity.
A very useful point is that this implies that the wave function has the same time dependence for all operators of a given generation $k$. This means we only have to compute a single representative $r$ for each $k$. This will make the computation of the wave function tractable.

Let's understand how this works by considering the expressions for two different generation $3$ operators, each  ``born'' from $| \includegraphics[scale = 1.3,valign=c]{gen2num1.pdf}\rangle$. The first expands ``depth-first,''
\be
\Big\langle \includegraphics[scale = 2,valign=c]{gen3num1.pdf}\Big|e^{-i\hat{H}t}\Big|\includegraphics[scale = 2,valign=c]{gen0.pdf}\Big\rangle \propto\int dt_1\, dt_2\, dt_3\, ~ \includegraphics[scale = 1.5,valign=c]{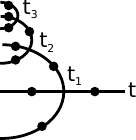} \qquad (0<t_3<t_2<t_1<t),\label{eq:gen-3-depth}
\ee
and the second expands ``breadth-first,''
\be
\Big\langle \includegraphics[scale = 2,valign=c]{gen3num2.pdf}\Big|e^{-i\hat{H}t}\Big|\includegraphics[scale = 2,valign=c]{gen0.pdf}\Big\rangle \propto\int dt_1\, dt_2\, dt_3\, ~ \includegraphics[scale = 1.5,valign=c]{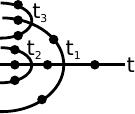} \qquad (0< t_3, t_2 < t_1 < t).\label{eq:gen-3-bredth}
\ee
We have not included constants of proportionality, because we will fix them below by a different argument. The propagators with a dot on them represent the dressed retarded propagators $G_R$, so that explicitly
\begin{align}
\includegraphics[scale = 1.5,valign=c]{gen3num1withtimes.pdf}~&=~ G_R(t_3)^{q-1}G_R(t_3-t_2)G_R(t_2)^{q-2}G_R(t_2-t_1)G_R(t_1)^{q-2}G_R(t_1-t), \label{eq:depth-first-GRs}\\ ~ \notag\\
\includegraphics[scale = 1.5,valign=c]{gen3num2withtimes.pdf}~&=~ G_R(t_3)^{q-1}G_R(t_2)^{q-1}G_R(t_3-t_1)G_R(t_2-t_1)G_R(t_1)^{q-3}G_R(t_1-t).\label{eq:breadth-first-GRs}
\end{align}
The ratio of these expressions is $G_R(t_{3} -t_2)G_R(t_1)/G_R(t_2)G_R(t_{3}-t_1)$. Since $G_R(t) = \theta(t)$ at leading order in large $q$, we see that the nontrivial time dependence of these integrands is equal. The only difference is that they have a different set of $\theta$ functions that impose different orderings of the time arguments. This applies more generally: we can write the integrand of any dressed fan diagram at generation $k$ in the simple form
\be
\prod_{j=1}^k \frac{1}{\cosh^{2} \CJ t_j} + O(1/q)\label{eq:large-q-integrand},
\ee
times a set of step functions that impose the ordering of time arguments appropriate for a given fan diagram.

We now have to do the integral. In principle, this integral should be over times $t_1 \dots t_k$ respecting the constraints from the step functions. In our example, the depth-first operator has $t_3<t_2<t_1$, and the breadth-first operator has $t_3 < t_1$ and $t_2 < t_1$, with no relationship between $t_2$ and $t_3$. However, since the integrand \eqref{eq:large-q-integrand} is symmetric under interchanges of the $t_j$, these restrictions will only affect the numerical prefactor and not the time dependence of the result. Thus, the time dependence of any dressed fan diagram will be the same. Picking the ``depth-first'' expansion to be our representative $r$ at each generation $k$, we have that
\be
\langle \mathcal{O}_k^{(r)}|e^{-i\hat Ht}|\includegraphics[scale = 1.3,valign=c]{gen0.pdf}\rangle \propto \bigg[ \int_0^{t}  \frac{dt_j}{\cosh^{2} \CJ t_j} \bigg]^{k}\propto \tanh^k \CJ t\label{eq:expression-for-wave-function-large-q}.
\ee
This implies that at leading order in large $q$, we have $P_{s_k}(t) \propto \tanh^{2k}(\mathcal{J} t)$, where $s_k = 1 + (q-2)k$.

As a final step, we need to determine the numerical coefficients. We can do this by requiring that the probability distribution remain normalized for all times. The trick here is to use the fact that we already have an expression for $P_{1}(t) = G_R(t)^2$ from \eqref{eq:q-retarded-propagators-large-q} that is accurate at first subleading order in the $1/q$ expansion:
\be
P_{1}(t) = 1 - \frac{4}{q} \log\cosh \CJ t + O(1/q^2).
\ee
Now, to determine the numerical coefficients for $P_{s_k}$ with $k>1$, we try to solve
\be
1 = P_{1}(t) + \sum_{k=1}^\infty \mathcal{N}_k \, \tanh^{2k} \CJ t,
\ee
to order $1/q$. Indeed, one can solve this equation by setting $\mathcal{N}_k = 2/kq$. This gives the probability distribution at leading nontrivial order in $1/q$
\be
P_s(t) = \Bigg\{
\begin{matrix*}[l] 
1-\frac{4}{q} \log\cosh \CJ t + O(1/q^2), &  s=1, \\
\frac{2}{kq} \tanh^{2k} \mathcal{J}t + O(1/q^2), & s= 1 + (q-2)k, & k = 1, 2, 3, \dots.
\end{matrix*}\label{eq:large-q-probability}
\ee
We will now make several comments about this result.

\begin{enumerate}
\item One can evaluate the expectation value of the size $s$ in this distribution. At leading order in $1/q$, we have:
\begin{align}\label{largeqsize}
\langle s\rangle &= \sum_k P_{s_k}s_k = 1 + \sum_{k =1}^\infty 2\tanh^{2k}\mathcal{J}t\\ &= \cosh(2\mathcal{J} t).
\end{align}
This result for the expected value of the size determines the initial exponential growth of the anticommutator-squared, via (\ref{anticommutator-squared}). We conclude that the chaos exponent at large $q$ and infinite temperature is $\lambda_L = 2\mathcal{J}$. The formulas from \cite{Maldacena:2016hyu} can be used to show that in the large-$q$ model we have $\lambda_L = 2\mathcal{J}\sqrt{1-x^2}$ where $x = \frac{q^2 E}{\mathcal{J}N}$. Here the energy spectrum is such that $-1<x<1$, and $x = 0$ corresponds to infinite temperature state. So we find agreement with previous results.

\item Note that the leading-order answer for (\ref{largeqsize}) depends on the $\frac{1}{q}$-suppressed probabilities for $s>1$, because $s \approx qk$ and this factor of $q$ cancels against the $\frac{1}{q}$ suppression. In other words, at large $q$, the operator initially has only small probability (of order $\frac{1}{q}$) to grow, but if it does grow it gets so big (size proportional to $q$) that this makes a large effect on the expected value of the size. This is reflected in the fractional variance of the size distribution, which is large, proportional to $q$
\be
\frac{\langle s^2 \rangle - \langle s \rangle^2 }{\langle s \rangle^2} = \frac{q}{2} \tanh^2 2 \CJ t + O(1).
\ee

\item As another example of something one can compute with this distribution, we can generalize the logic that led to (\ref{anticommutator-squared}) slightly, finding
\be
\frac{1}{N}\sum_{j}\frac{1}{2^{N/2}}\text{Tr}\left[\{\psi_1(t),\psi_j\}\{\psi_1(t'),\psi_j\}\right] = \sum_s \frac{s}{N}\sqrt{P_s(t)P_s(t')}.
\ee
Evaluating this with our large $q$ result (\ref{eq:large-q-probability}), we find
\be
\frac{1}{N}\sum_{j}\frac{1}{2^{N/2}}\text{Tr}\left[\{\psi_1(t),\psi_j\}\{\psi_1(t'),\psi_j\}\right] = \frac{1}{N}\frac{\cosh \left[\mathcal{J}(t+t')\right]}{\cosh \left[\mathcal{J}(t-t')\right]} + O(\frac{1}{Nq}) + O(\frac{1}{N^2}).
\ee

\item So far, we have considered a simple large-$q$ limit, where we do not allow $t$ to scale with $q$. This approximation will break down at times of order $q$. It would be nice to extend our analysis to resum effects of order $t/q$. Although we have not studied this systematically, we will make a few comments. To capture the important effects, one can no longer approximate $G_R(t)$ as simply $\theta(t)$ in cases where the time argument can be long. Instead, we can approximate it as $G_R(t) = \theta(t)\cosh^{-2/q}(\mathcal{J} t) \approx \theta(t)e^{-2\mathcal{J}t/q}$. We expect based on numerics and \cite{Tarnopolsky:2018env} that this expression is accurate for all time $t$, although it does not follow from the approximation worked out in \cite{Maldacena:2016hyu}.

A convenient feature of this approximation is that (ignoring the step functions) we have $G_R(t)G_R(t') = G_R(t+t')$. This composition property allows us to convert fan diagram integrands into each other, so we retain the property that only one representative from each generation must be computed. In the example given above, to convert (\ref{eq:depth-first-GRs}) to (\ref{eq:breadth-first-GRs}), we use $G_R(t_3-t_2)G_R(t_1) = G_R(t_3-t_1)G_R(t_2)$. Another simplification is that inside the integrand, we can expect to approximate $G_R(t_j)^{q-\alpha} \approx G_R(t_j)^q$, because the presence of the factor $G_R(t_j)^q$ will make the integral prefer the region where $t_j$ is of order one, so the factor $G_R(t_j)^\alpha$ will be approximately one.

Following these approximations, we find that the total effect is to multiply the wave function (\ref{eq:expression-for-wave-function-large-q}) by $G_R(t)$. Normalizing the probability distribution, we find the expression
\be
P_s(t) = \frac{\Gamma(k + 2/q)}{\Gamma(k+1)\Gamma(2/q)}\frac{\tanh^{2k}\mathcal J t}{\cosh^{4/q}\mathcal{J}t}, \hspace{20pt} s = 1 + (q-2)k.\label{resummed}
\ee
This probability distribution resums the $t/q$ corrections to our straightforward large-$q$ result (\ref{eq:large-q-probability}). However, it is not fully satisfactory because there are expected to be $k/q$ corrections that are not accurately summed here. We hope that the expression is nevertheless qualitatively accurate even for large $k$. Note that at our level of approximation the denominator could be written $e^{4\mathcal{J}t/q}$ instead of $\cosh^{4/q}\mathcal{J}t$.

Our main purpose in writing the expression (\ref{resummed}) is that one finds a very similar formula in a classical model of operator growth discussed in \S~\ref{appendixB}.

\item It is sometimes convenient to define a ``coarse-grained'' wave function by $\Psi_s(t) = (-i)^k\sqrt{P_s(t)}$. This is the amplitude for $\psi_1(t)$ to be of size $s$ at time $t$. We ought to have a composition property where the two point function of fermions can be computed by inserting a complete set of states at an intermediate time and summing over the sizes of all operators that appear. In other words, we should have
\begin{align}
\Psi_1(t_1+t_2) &= \langle \psi_1(t_1+t_2)|\psi_1(0)\rangle = \sum_{s}\langle \psi_1(t_1+t_2)|\mathcal{O}_s(t_2)\rangle\langle \mathcal{O}_s(t_2)|\psi_1(0)\rangle,\notag\\&= \sum_{s} \Psi^*_s(t_1)\Psi_s(-t_2).
\end{align}
Indeed, one can check that this property holds for (\ref{resummed}). It follows that it also holds at order $\frac{1}{q}$ for (\ref{eq:large-q-probability}).
\end{enumerate}

\section{Discussion}\label{sec:discussion}

In this paper, we discussed the time evolution of a simple fundamental fermion operator $\psi_1$ in the SYK model. In the large $N$ limit, we related the operator growth problem to the problem of a particle moving on a rapidly expanding graph. We computed the size distribution for the evolving operator explicitly in two cases: numerically for $N=30$ fermions with $q=4$ and analytically at large $N$ and large $q$. We showed how to use this size distribution to compute out-of-time-order correlators at infinite temperature.

Throughout, we have emphasized a particular decomposition of the time evolving operator, into components with a given size (number of elementary fermions appearing in a product). We would like to contrast this with the Baker-Campbell-Hausdorff expansion
\be
\psi_1(t) = e^{iHt}\psi_1 e^{-iHt} = \psi_1 +it[H, \psi_1 ] - \frac{t^2}{2}[H,[H, \psi_1]] -  \frac{it^3}{3!}[H,[H,[H, \psi_1]]] + \dots.  \label{eq:BCH}
\ee
These terms also give a decomposition of the time-evolving operator. We emphasize that this is different from the size decomposition, because the terms at order $k$ in the BCH expansion do not all have the same size.
While the $k$th nested commutator contains terms up to size $s = k(q-2) + 1$, there is also weight on operators of shorter sizes. For instance, in the $q=4$ SYK model, the $k=2$ term in the BCH expansion contains operators of size $5$ as well as an operator of size $1$, namely $\psi_1$. The fact that our wave function $\sqrt{P_s(t)}$ is a nontrivial function of time indicates that it receives contributions from many different orders in the BCH expansion (starting at order $k$, where $s = 1 + (q-2)k$).

Another point we would like to emphasize is the following. The notion of size that we have used makes explicit reference to a particular set of simple operators $\{\psi_i\}$, out of which we construct complicated ones. This set of simple operators is determined by the Hamiltonian itself, and it depends on the $q$-local and sparse nature of $H$. If instead the Hamiltonian were a totally random matrix, we would have no sensible notion of simple operators, and no good way to define size. However, for Hamiltonians such as SYK, a preferred set of simple operators is selected by the fact that the interaction can be written in terms of finite (order $q$) products of them.

There are many possible directions for improvements on our work. For example, it would be interesting to understand $1/N$ corrections to the $P_s(t)$ distribution at a level that would make it possible to see saturation of the late-time distribution. Another challenge is to extend the approach studied here to compute out-of-time-order correlators at finite temperature. For instance, one might be tempted to try to define a size distribution $P_s^{(\beta)}(t)$ with respect to an inner product $(A,B)_\beta \equiv Z(\beta)^{-1}\, \Tr[A^\dagger \, e^{-\beta H /2} \, B \, e^{-\beta H /2}]$, where $Z(\beta)$ is the thermal partition function. In principle, one could use the Schwinger-Dyson equations at large $q$ to compute a candidate wave function. However, this necessarily requires the use of a different set of operators $\CO_k^{(\ell)}(\beta)$ that now depend on the temperature $\beta$.  Unfortunately these operators do not appear to admit a simple relationship between ``generation'' $k$ and operator size $s$. This means we do not know how to extract the expected size from this candidate distribution, and we do not know how to relate it to the out-of-time-order commutator as we did in \S~\ref{sec:size-and-com}.

In holographic theories, operator growth is described by a particle falling towards the black hole horizon. It is tempting to think of the radial direction in the graph of operators as being similar to the radial direction in the bulk theory, so that the particle propagating deeper into the graph resembles the particle falling into the bulk. We do not know if there is a more precise connection to be made there.

\section*{Acknowledgments}
We are grateful to Juan Maldacena, Xiao-Liang Qi, Steve Shenker, Lenny Susskind, and Beni Yoshida for discussions. 
DR acknowledges support from the Simons Foundation through the ``It
From Qubit'' collaboration as well as funding from the Paul Dirac Fund of the Institute for Advanced Study and the NSF grant PHY-1314311. DR would also like to acknowledge the Aspen Center for Physics, which is supported by National Science Foundation grant PHY-1607761, and the Kavli Institute for Theoretical Physics, which is supported in part by the National Science Foundation under Grant No. NSF PHY11-25915.
DS is supported by Simons Foundation grant 385600, and AS is supported by the Simons Foundation. 
This paper has been brought to you graphically by the letter $\psi$.

\appendix

\section{Some numerical data}\label{sec:numerical-data}
In this appendix we give numerical data for the infinite-temperature chaos exponent $\lambda_L$ and the rate of decay of the two point function in real time, $\mu$, for different values of $q$. The parameter $\mu$ is defined by saying that for large $t$, $G(t)\propto e^{-\mu t}$.

Although we do not know how to compute these quantities analytically, it is straightforward to compute them by numerically solving the Schwinger-Dyson equations in real time. This directly gives $\mu$, and by constructing the retarded ladder kernel \cite{kitaevfirsttalk}, one can find $\lambda_L$ as described in \cite{Maldacena:2016hyu}. We give a table here of the values that we found.\begin{table}[h]
\footnotesize
\begin{center}
\begin{tabular}{c||c|c|c|c|c|c|c|c|c|c|c|c|c}
q&2&2.1&2.2&2.5&3&4&5.5&7&10&15&20&30&50\\
\hline
$\frac{\lambda_L}{2\mathcal{J}}$&0&0.08&0.15&0.30&0.454&0.620&0.738&0.799&0.863&0.910&0.934&0.956&0.974\\
\hline
$\frac{\mu}{2\mathcal{J}}$&0&0.10&0.17&0.29&0.43&0.446&0.318&0.202&0.123&0.0756&0.0548&0.0353&0.0207
\end{tabular}
\caption{{\small Values for the infinite temperature chaos exponent $\lambda_L$ and the rate of decay of the infinite temperature two point functions $\mu$. Note that the values for smaller values of $q$ are less precise.}}
\end{center}
\end{table} Note that although the physical model makes sense only for even integer $q$, the Schwinger-Dyson equations make sense for arbitrary $q$. Note that the values of $\mu$ for fairly large $q$ seem to agree well with the formula $\frac{\mu}{2\mathcal{J}} = \frac{1}{q} + \frac{\pi^2}{6 q^2}+O(q^{-3})$ that one would expect based on \cite{Tarnopolsky:2018env}.

\mciteSetMidEndSepPunct{}{\ifmciteBstWouldAddEndPunct.\else\fi}{\relax}
\bibliographystyle{utphys}
\bibliography{operator}
\end{document}